\documentclass[aps,twocolumn,showpacs,prl]{revtex4}     

\usepackage{graphicx}                   
\usepackage{epsfig}
\usepackage{setspace}

\begin{document}


\title{Theoretical phase diagram of ultrathin films of
incipient ferroelectrics}

\author{A. R. Akbarzadeh$^{1,2}$, L. Bellaiche$^{2}$, Jorge
\'I\~niguez$^{3}$, and David Vanderbilt$^{4}$}

\address{$^{1}$Department of Materials Science and Engineering,
University of California, Los Angeles, P.O. Box 951595, Los
Angeles, California 90095-1595, USA}
\address{$^{2}$Physics Department, University of Arkansas,
Fayetteville, AR 72701, Arkansas, USA}
\address{$^{3}$Institut de Ciencia de Materials de Barcelona
(ICMAB-CSIC), Campus UAB, 08193 Bellaterra, Spain}
\address{$^{4}$Department of Physics and Astronomy, Rutgers
University, Piscataway, New Jersey 08854-8019, USA}

\date{\today}


\begin{abstract}
We have used a first-principles-based scheme to compute the
temperature-versus-misfit strain ``Pertsev'' phase diagram of
ultrathin films of {\sl incipient} ferroelectric KTaO$_3$. Our results
suggest that, at variance with the bulk material, KTaO$_3$ ultrathin
films cannot be described as quantum paraelectrics. Rather, the
behavior of the films is largely determined by surface/interface
effects that favor ferroelectricity and the imperfect screening of the
depolarizing fields. This leads to Pertsev phase diagrams that are
qualitatively similar to those of {\sl normal} ferroelectrics such as
BaTiO$_3$.
\end{abstract}



\narrowtext

\maketitle

\marginparwidth 2.7in \marginparsep 0.5in

The need for miniaturized devices and the quest for knowledge in
nanoscience have led to a flurry of activities in ferroelectric thin
films  that deepened our understanding of such low-dimensional systems
(see Refs.~\onlinecite{Dawber,ScottScience,Igorreview} and references
therein).  In particular, the crucial influence of surface/interface
effects and mechanical and electrical boundary conditions on the
properties of ferroelectric thin films is now well documented.  For
instance, recent studies revealed that the strain arising from the
substrate can lead to ferroelectric phases absent in the
bulk~\cite{Bo,pertsev,dieguez212101} and that surface/interface
effects can yield an asymmetric temperature-versus-misfit strain
diagram~\cite{Bo,rios}. Similarly, residual depolarizing fields are
now known to generate anomalous effects in ferroelectric ultrathin
films, such as the existence of stripe domains with remarkably small
periods~\cite{Streiffer,IgorPRL,Inna}.

An important issue that remains to be fully understood concerns the
properties of ferroelectric films made of the so-called incipient
ferroelectrics, i.e., materials such as SrTiO$_3$ and KTaO$_3$ for
which the bulk phase ferroelectricity is suppressed by quantum
fluctuations~\cite{viana94,KTa04,slace94}. The pioneering work in
Ref.~\onlinecite{Haeni:Nature758} revealed the existence of
room-temperature ferroelectricity in strained 500\,\AA-thick SrTiO$_3$
films, and can thus be taken to suggest that the coupling between
strain and dipoles tends to prevail over quantum effects in
nanostructures. One may thus wonder what the effect (if any) of
quantum zero-point vibrations is in ultrathin films made of incipient
ferroelectrics. In particular, the following questions are, to the
best of our knowledge, currently unanswered: (1) Can quantum
fluctuations overcome the coupling between strain and dipoles in
epitaxially {\it strained} films and, for example, suppress some
phases? If so, which phases? (2) For zero misfit strain, are the films
paraelectric or ferroelectric? (3) What does the coexistence of
quantum and surface/interface effects lead to? (4) What are the
consequences and signatures of the zero-point motion of the ions in a
film experiencing a depolarizing field?

The aim of this Letter is to answer these questions by determining and
analyzing the temperature-versus-misfit strain phase diagrams of
KTaO$_3$ ultrathin films under different electrical boundary
conditions, using effective models of the films derived from
first-principles. As we will see, many interesting features will be
revealed and explained.

First-principles-based effective Hamiltonian methods have been very
fruitfully employed to study ferroelectric thin films before (see,
e.g., Refs.~\onlinecite{Bo,IgorPRL,Inna}, and
\onlinecite{Ghosez}). Here we used such an approach to study KTaO$_3$
ultrathin films grown along the [001] pseudo-cubic direction (chosen
to be along the z-axis), having K-O terminated surfaces/interfaces,
and being 28\,\AA\ thick. Technically, the films are modeled by
10$\times$10$\times$7 supercells that are periodic along the x- and
y-axes (which lie along the [100] and [010] pseudo-cubic directions,
respectively) and contain seven TaO$_2$ (001) layers stacked along the
non-periodic z-axis. The total energy of such a supercell is written
as
\begin{equation}\label{eq:eq1}
\mathcal{E}_{\rm tot}=\mathcal{E}_{\rm Heff}({\mathbf u_{\rm i}},{\mathbf v_{\rm i}},\eta)+
   \beta \mathbf{E_{\rm d}} \cdot \sum_{\rm i} Z^* \mathbf{u_{\rm i}} \;,
\label{eq:1}
\end{equation}
where ${\mathbf u_{\rm i}}$ is the local soft mode in the unit cell {\rm i} of
the film, such that $Z^* \mathbf{u_{\rm i}}$ yields the local
electrical dipole, and $\beta \mathbf{E_{\rm d}}$ is a term related to
screening of the depolarization field, as discussed below. 
The ${\bf \{ v_{\rm i}\}}$'s variables describe
inhomogeneous strains within the supercell~\cite{zhon95}. $\eta$ is
the homogeneous strain tensor, which is particularly relevant to
mechanical boundary conditions since epitaxial (001) films are
associated with the freezing of some components of $\eta$ (in Voigt
notation), i.e., $\eta_6$\,=\,0 and $\eta_1$\,=\,$\eta_2$\,=\,$\delta$, with
$\delta$ being the value forcing the film to adopt the in-plane
lattice constant of the
substrate~\cite{pertsev,dieguez212101,IgorPRL}. In practice,
$\delta$=$(a_{\rm sub}-a_{\rm KTa})/a_{\rm KTa}$, where $a_{\rm sub}$
is the in-plane lattice parameter of the substrate and $a_{\rm
KTa}$\,=\,3.983\,\AA\ is the 0\,K cubic lattice constant of bulk
KTaO$_3$ used in Ref.~\onlinecite{KTa04}. 

The expression for $\mathcal{E}_{\rm Heff}$, the intrinsic
effective-Hamiltonian energy of the film, is given in
Ref.~\onlinecite{zhon95}. The first-principles-derived parameters for
bulk KTaO$_3$ are those of Ref.~\onlinecite{KTa04}, except that here
we use the formula for the dipole-dipole interactions for thin films
under ideal open-circuit (OC) boundary conditions derived in
Refs.~\onlinecite{InnaPRB} and \onlinecite{Ivancondmat}. Such
electrical boundary conditions naturally lead to the existence of a
maximum depolarizing field inside the film (denoted by $\mathbf E_{\rm
d}$) when the dipoles all point along the [001] direction. The second
term of Eq.~(1) mimics a screening of $\mathbf E_{\rm d}$ via the
parameter $\beta$. More precisely, the residual depolarizing field
resulting from the combination of the first and second terms of
Eq.~(1) has a magnitude equal to $(1-\beta)\,\mathbf E_{\rm d}$. As a
result, $\beta$\,=\,0 corresponds to ideal OC conditions, an increase
$\beta$ lowers the magnitude of the resulting depolarizing field, and
$\beta$\,=\,1 corresponds to ideal short-circuit (SC) conditions for which
the depolarizing field vanishes. Note that $\mathbf E_{\rm d}$ depends
on the dipole configuration, is exactly derived at an atomistic level
(following the procedure introduced in Ref.~\onlinecite{InnaPRB}), and
is self-consistently updated during the simulations.

We should note here the main approximations made in the construction
of the above described model for ultrathin films of KTaO$_3$ under
varying electrical and mechanical boundary conditions. Firstly, as in
previous works~\cite{Bo,Ghosez}, our $\mathcal{E}_{\rm tot}$ relies on
a simple truncation, at the surface/interface layers, of the
interactions existing in bulk KTaO$_3$.  While this approximation is
admittedly crude, the resulting $\mathcal{E}_{\rm tot}$ mimics in a
qualitatively correct way the enhancement of the surface/interface
polar modes that occurs in the most promising compounds (see, e.g.,
the experimental work on Ba$_{0.5}$Sr$_{0.5}$TiO$_3$ films in
Ref.~\onlinecite{rios} or the {\sl ab initio} results for PbTiO$_3$ films sandwiched
by Pt electrodes in Ref.~\onlinecite{sai}). In our simple model,
such an enhancement happens because our truncation removes short-range
interactions that oppose the onset of such local dipoles. At a
quantitative level, this approximation may well be underestimating the enhancement for some
choices of electrodes and overestimating it for others. At any rate,
as the interesting effects resulting from our simulations are very
pronounced, we are confident they can be taken as reliable qualitative
predictions.

Secondly, we made calculations for ideal electrodes leading to SC
boundary conditons (i.e., $\beta$\,=\,1), and also for realistic
situations for which an appropriate different value of $\beta$ must be
chosen. Let us mention two pieces of information to justify such a
choice: (i) the {\sl ab intio} studies of Sai {\sl et al.}~\cite{sai},
who found a 97\,\% cancellation of the depolarizing field in some
systems, and (ii) ongoing unpublished work of one of our co-authors
(L.~Bellaiche) who, by comparing measurement and simulations, has
estimated that some experimental set-ups correspond to a screening of about 98\,\% of the
maximum depolarizing field. Here, we made a conservative choice and used
$\beta$\,=\,0.96. We thus expect the striking effects predicted will be robust as far as this 
approximation is concerned.

As in Ref.~\onlinecite{KTa04}, $\mathcal{E}_{\rm tot}$ is used in two
different kinds of Monte-Carlo (MC) simulations: classical Monte Carlo
(CMC)~\cite{Metropolis}, which does not take into account zero-point
vibrations, and path-integral quantum Monte Carlo
(PI-QMC)~\cite{zhong,jorge,Ceperely95}, which includes the
quantum-mechanical zero-point motions.  Consequently, comparing the
results of these two different Monte-Carlo techniques allows a precise
determination of quantum effects on properties of the considered
KTaO$_3$ ultrathin films. We typically used 30,000 MC sweeps to
thermalize the system and 70,000 more to compute averages, except at
low temperature in PI-QMC where more statistics are needed.  Further
details about the PI-QMC technique may be found in
Refs.~\onlinecite{jorge,KTa04}, and \onlinecite{Ceperely95}.
%
\begin{figure}[htb]
\includegraphics[height=0.3\textheight, width=0.5\textwidth]{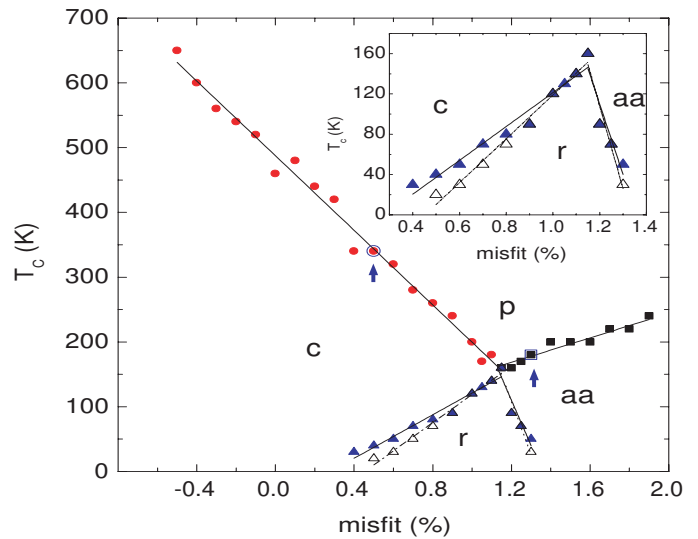}
\caption{Temperature \emph{versus} misfit strain diagram of a
28\,\AA-thick KTaO$_3$ ultrathin film under ideal SC conditions
($\beta=1$). Solid and open symbols refer to CMC and PI-QMC
simulations, respectively. Serving as guides for the eyes, the
solid and dashed lines are linear fits of the CMC and PI-QMC data,
respectively.  The inset shows a magnified view of the region in which
the {\sl r} phase occurs.}\label{f:Fig1}
\end{figure}

Figure~1 displays the ``Pertsev'' (that is, the temperature
\emph{versus} misfit strain $\delta$) phase diagram for the
28\,\AA-thick KTaO$_3$ thin film under ideal SC conditions, as
resulting from CMC (solid symbols) and PI-QMC (open symbols)
simulations.  One can clearly see that both kinds of simulations
generate the four phases that also appear in the Pertsev diagram of
(001) BaTiO$_3$ ultrathin films~\cite{Bo,dieguez212101}. These phases
are: a paraelectric {\sl p} state at high temperatures; a
ferroelectric tetragonal {\sl c} phase for intermediate temperatures
and compressive and weakly tensile strains, having a polarization
lying along the [001] growth direction; a ferroelectric orthorhombic
{\sl aa} phase for low temperatures and large tensile strains, in
which the polarization is parallel to the in-plane [110] direction;
and a ferroelectric monoclinic {\sl r} phase for the lowest
temperatures and intermediate tensile strains, for which the
polarization direction continuously rotates from [001] to [110] as
$\delta$ increases.

Figure~1 further displays four remarkable features. (1) Quantum
effects begin to appear at temperatures below $\approx$\,100\,K (we
numerically found that the transition lines predicted by CMC and
PI-QMC overlap above 100\,K), as in bullk KTaO$_3$~\cite{KTa04}. As a
result, and as emphasized by the inset of the figure, the {\sl
c}--to--{\sl r} and, to a lesser extent, the {\sl aa}--to--{\sl r}
phase boundaries are the {\it only} boundaries affected by quantum
fluctuations (PI-QMC simulations performed at temperatures higher than
100\,K, at 0.5\,\% and 1.3\,\% misfit tensile strains, are indicated
with arrows in Fig.~1, and reveal that the high-temperature {\sl
p}--to-{\sl c} and {\sl p}--to-{\sl aa} phase transitions are
unaffected by quantum fluctuations). The effects are, in any case,
minor. Essentially, there is a small reduction of the strain range in
which the {\sl r} phase occurs: in the low-temperature limit, it
passes from about +0.4\,\%$\leq\,\delta\,\leq$\,+1.3\,\% at a
classical level to about +0.5\,\%\,$\leq\,\delta\,\leq$\,+1.3\,\% when
quantum fluctuations are included. (2) As in BaTiO$_3$ ultrathin
films~\cite{Bo,dieguez212101}, the {\sl p}, {\sl c}, {\sl aa}, and
{\sl r} phases {\sl meet} at a single four-phase point, which occurs
here at a temperature of about 160\,K and a {\it tensile} strain of
about 1.12\,\% for both CMC and PI-QMC simulations~\cite{footnote1}.
(3) The phase diagram is asymmetric with respect to zero misfit
strain.  Such an asymmetry is hinted in experiments on
Ba$_{0.5}$Sr$_{0.5}$TiO$_3$ films~\cite{rios} and is 
related to the enhancement of the z-component of the local dipoles
at the surfaces/interfaces. As mentioned above, in our simulations the
enhancement occurs because the polar local modes at the
surfaces/interfaces are partly free from energetically-costly
short-range interactions~\cite{footnote2}. (4) For zero misfit strain, the
paraelectric--to--\emph{c} transition occurs at about 460\,K for both
CMC and PI-QMC simulations.

The high transition temperature ($T_{\rm C}$) obtained for the film at zero
misfit strain starkly contrasts with the behavior the bulk material,
for which the paraelectric--to--ferroelectric transition occurs at
about 30\,K at a classical level and vanishes when quantum effects are
considered~\cite{KTa04}. The physical origin of such a high $T_{\rm C}$ lies
on the above mentioned dipole enhancement at the surface/interface of
the films. The striking consequence of such an enhancement is that
these ultrathin films are no longer incipient ferroelectrics, but
display a Pertsev phase diagram that is in essence identical to that
of a normal ferroelectric like BaTiO$_3$.

It is also interesting to realize that SrTiO$_3$ has a lattice
constant about 2\,\% smaller than KTaO$_3$, and that a linear
extrapolation of the data in Fig.~1 suggests that growing KTaO$_3$
ultrathin films on a SrTiO$_3$ substrate (for which
$\delta\,=\,-2$\,\%) should lead to a $T_{\rm C}$ that could be as
high as 1000\,K. In other words, our simulations suggest that growing
an ultrathin film made of an incipient material (that is, KTaO$_3$) 
on a substrate made from another incipient ferroelectric (namely, SrTiO$_3$) 
may result in a ferroelectric compound with a high
$T_{\rm C}$ and, thus, a large polarization {\sl along the growth
direction} at low temperatures. This would be the result of, not only
surface/interface effects, but also the well-known coupling between
dipoles and compressive strain~\cite{pertsev2000,king94,cohen92}.
%
\begin{figure}[htb]
\includegraphics[height=0.3\textheight, width=0.5\textwidth]{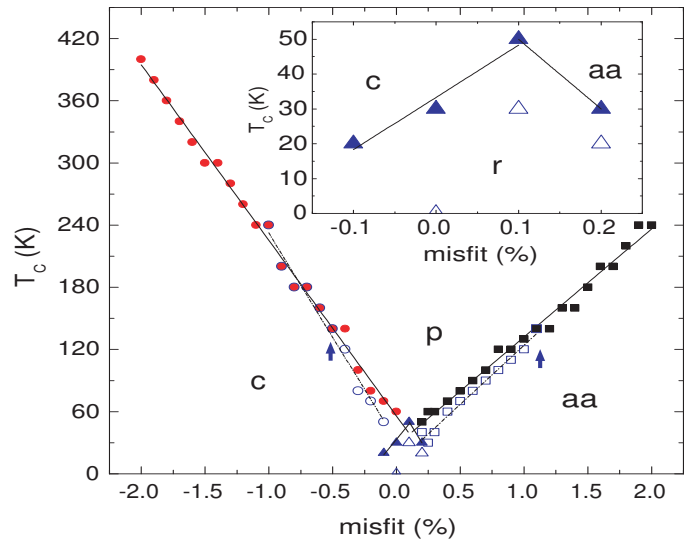}
\caption{Temperature \emph{versus} misfit strain diagram of a
28\,\AA-thick ultrathin film KTaO$_3$ under a residual depolarizing
field ($\beta$=0.96). Solid and open symbols refer to CMC and
PI-QMC simulations, respectively.  The solid and dashed lines
are linear fits of the CMC and PI-QMC data, respectively, and are
guides for the eyes. The inset is a magnification of the region
in which the {\sl r} phase occurs, with the PI-QMC data point for
the {\sl c}--to--{\sl r} transition temperature at zero misfit strain
being a guess resulting from the interpolation of the PI-QMC
simulations for the lowest computed temperatures.}\label{f:Fig2}
\end{figure}

We now discuss how depolarizing fields, always present in reality as, e.g., 
the metallic electrodes are never ideal \cite{JavierNature}, affect the Pertsev phase
diagram of KTaO$_3$ ultrathin films. As explained above, we have
chosen $\beta$\,=\,0.96 in Eq.~(1), i.e., a situation that corresponds to a screening of 96\,\% of
the maximum depolarizing field. The resulting Pertsev diagram is shown in
Fig.~2, for both CMC and PI-QMC simulations. By comparing the CMC
results from Figs.~1 and 2, it is clear that the depolarizing field
tends to suppress the z-component of the polarization, as expected.
Indeed, this effect dramatically decreases the {\sl p}--to--{\sl c}
transition temperatures; for instance, the transition occurs around
60\,K at zero misfit strain, i.e., the $T_{\rm C}$ is 400\,K lower
than the one obtained for ideal SC. It also considerably extends the
region of small tensile strain associated with the $aa$ phase, which
exhibits an in-plane polarization that does not directly feel the
depolarizing field, and significantly narrows the {\sl r-phase}
region.

As a result, the phase diagram becomes more symmetric around the zero
misfit strain and the four-phase point of Fig.~1 shifts towards lower
temperature (about 50\,K for CMC simulations~\cite{footnote1}).  As
shown in Fig.~2, such a small transition temperature implies that the
quantum fluctuations (which are appreciable only at temperatures below
120\,K, and become more pronounced as the temperature decreases) have
two remarkable effects that do {\it not} occur in the film under SC
conditions.  First of all, the zero-point vibrations now affect the
{\sl p}--to--{\sl c} and {\sl p}--to--{\sl aa} phase boundaries (for
small strain) -- in addition to the boundaries involving the {\sl r}
phase. Secondly, quantum fluctuations significantly reduce the
stability range of the {\sl r-phase}. This suggests that, for slightly
worse electrodes (i.e., for $\beta\,<$\,0.96), quantum effects could
well lead to the disappearance of the four-phase point in favor of a
{\it three}-phase point in which ``only'' the {\sl p}, {\sl c}, and
{\sl aa} meet.

In summary, we have used a first-principles-based scheme to determine
the temperature-versus-misfit strain ``Pertsev'' phase diagram of
ultrathin films of {\sl incipient} ferroelectric KTaO$_3$ under
short-circuit-like boundary conditions.  We have performed both
classical and quantum Monte Carlo simulations, the ionic quantum
fluctuations being fully taken into account in the latter. This has
allowed us to investigate the competition of the various effects
present in the films, namely, the misfit strains and surface/interface
effects that are known to favor the occurrence of ferroelectric
phases, and the zero-point vibrations and residual depolarizing fields
that tend to suppress ferroelectricity.

We have found that, for ideal electrodes, quantum fluctuations have a
negligible effect; the phase diagram is essentially determined by the
other factors mentioned above and closely resembles what is found in,
e.g., BaTiO$_3$ films. Our results thus suggest that, in ultrathin
film form, {\sl quantum paraelectrics} such as KTaO$_3$ behave
qualitatively in the same way as {\sl strong} ferroelectrics such as
BaTiO$_3$. We have also simulated realistic (thus imperfect)
electrodes and found that the corresponding depolarizing fields have a
great impact on the calculated phase diagram; in particular, the
transition temperatures of the {\sl c} and {\sl r} phases are
significantly reduced. As a result, quantum mechanical effects become
more important in this case, and may alter the phase diagram
qualitatively, e.g., by suppressing the {\sl r} phase.

We believe the present study leads to a better understanding of
ferroelectric thin films, and hope it will stimulate experimental work
aimed at confirming our predictions.

We thank George A. Samara 
and Kevin Leung for useful discussions.  This work
is supported by ONR grants N00014-01-1-0365, N00014-04-1-0413,
 and N00014-05-1-0054, by NSF grant DMR-0404335,
and by DOE grant DE-FG02-05ER46188. J.I. thanks support from the
Spanish Ministry of Science and Education (FIS2006-12117-C04-01)
and FAME-NoE.

%



\begin{thebibliography}{99}
\bibitem{footnote2} For zero misfit strain, $\beta$\,=\,1, and
  $T$\,=\,50\,K, both CMC and PI-QMC rendered ${\rm u_{\rm
  z}}\, \approx$\,0.041 and 0.037~a.u. for surface/interface and inner
  layers, respectively. For $\beta$\,=\,0.96, these values get reduced
  to 0.015 and 0.013~a.u., respectively, for CMC, and to 0.013 and 0.012~a.u., 
  respectively, for PI-QMC. The corresponding magnitude of the local mode 
  is 0.024~a.u. at $T$\,=\,20\,K in bulk KTaO$_{3}$, according to CMC.
\end{thebibliography}
\end{document}